\documentclass[11pt,a4paper]{article}

\usepackage{amssymb}
\usepackage{graphicx}
\usepackage{subfigure}  %side-by-side figures
\usepackage{amsmath}
\usepackage{verbatim}
\usepackage{multirow}
\usepackage{algorithmic}
\usepackage[]{algorithm2e}
\usepackage{graphicx}
\usepackage{tikz}
\usepackage{fullpage}

\newcommand{\multiPicWidth}{0.45}%

\newcommand{\thrLow}{1~Mbps}%
\newcommand{\thrMed}{1.5~Mbps}%
\newcommand{\thrHigh}{2.6~Mbps}%

\title{A Novel Play-out Algorithm for HTTP Adaptive Streaming}
\author{Arkadiusz Biernacki\\
	Institute of Computer Science\\
Silesian University of Technology\\
Akademicka 16\\
44-100 Gliwice, Poland\\
E-mail: arkadiusz.biernacki@polsl.pl}

\begin{document}

\maketitle
\begin{abstract}
In the paper, we proposed a novel algorithm dedicated to adaptive video streaming based on HTTP. The algorithm employs a hybrid play-out strategy which combines two popular approaches: an estimation of network bandwidth and a control of a player buffer. The proposed algorithm was implemented in two versions which differ in the method of handling fluctuations of network throughput. 

The proposed hybrid algorithm was evaluated against solutions which base their play-out strategy purely on bandwidth or buffer level assessment. The comparison was performed in an environment which emulated two systems: a Wi-Fi network with a single immobile node and HSPA (High Speed Packet Access) network with a mobile node. The evaluation shows that the hybrid approach in most cases achieves better results compared to its competitors, being able to stream the video more smoothly without unnecessary bit-rate switches. However, in certain network conditions, this score is traded for a worse throughput utilisation compared to other play-out strategies.
\end{abstract}


\section{Introduction}
During the past years, web based video sharing services like YouTube, Hulu or Dailymotion have become very popular. The users of YouTube, which allows for the distribution of user-produced multimedia content, alone request millions of videos every day \cite{_youtube_????}. Consequently, popularity of this kind results in a drastic shift in Internet traffic statistic, which reports increase in traffic from Web-based video sharing services. It is estimated that traffic will account for about 80\% to 90\% of the global Internet traffic in a next few years, according to the recent report published by Cisco \cite{cisco_global_2013}.

Video streaming in the above mentioned services is either web-based or HTTP-based; therefore, being transported using TCP. HTTP and TCP are general purpose protocols and were not primarily designed for streaming of multimedia. Thus, attempts are being made to adapt delivery of multimedia content to the Internet environment. One of such attempts tries to introduce an additional layer of application control to transmitted video traffic. Since TCP is designed to deliver data at the highest available transmission rate, it may sometimes be reasonable for a sender to provide additional flow control if it is not strictly necessary for application data to reach a receiver as fast as the TCP would otherwise allow. Therefore, an application may limit the rate at which the data is transmitted, and, if the video bit-rate is lower than the end-to-end available bandwidth, the traffic characteristic will not resemble the characteristics of a standard TCP flow. Furthermore, modern video players implement stream-switching (or multi bit-rate): the content, which is stored at the web server, is encoded at different bit-rate levels, then an adaptation algorithm selects the video level, which is to be streamed, based on a state of a network environment or on a state of a video player. 

When the judgment is based on the state of the network environment, the video client estimates how fast the server can deliver video (i.e. the available capacity), e.g. by measuring arrival rate of video data. Then the client choses the video bit-rate which corresponds to the estimated network throughput. If it selects a video bit-rate that is too high, the viewer will experience re-buffering events, i.e. a playback will be suspended because the data transmission will not keep pace with a video bit-rate. If it picks a video bit-rate that is too low, a viewer will experience suboptimal video quality and part of the network bandwidth will be wasted.

When the decision is based on the state of the video player, the algorithm tries directly observe and control the playback buffer instead of estimating network capacity, believing that the buffer occupancy contains a lot of information and is a controllable variable. It is assumed that the buffer occupancy reflects the end-to-end system capacity, including current load conditions of the network and its rate of change reflects the mismatch between the network throughput and the requested video bit-rate.

In this work, we propose a new hybrid algorithm which combines the two above approaches: the bandwidth estimation and the buffer control. We assume that the hybrid solution will exploit strengths of both approaches and it will avoid their weaknesses. We compare performance of the hybrid solution with the two above described approaches subjecting them to variable network throughput, which is commonly encountered in wireless and mobile networks. During the performance evaluation of the algorithms, we measure how often the player buffer runs out, how long it takes to re-start the video transmission, how efficiently the algorithm utilises available network throughput, and finally, how often the player switches between different video rates.

We conduct this performance study using an emulation model. The emulation approach allows us to methodologically explore the behaviour of the examined system over a wide range of parameter settings, which would be a challenging task if we conducted such experiments only on a real-network. Simultaneously, as the emulation is performed in a laboratory environment, we are able to preserve much of the network realism because we conduct experiments using real hardware and software, which allows us to maintain a high level of accuracy for the obtained results.

\section{Theoretical background}
\subsection{Application level flow control}
\label{sec:appFlowControl}
One of the popular video transmission method is a progressive download, which is simply a transfer of a video file from an HTTP server to a client where the client may begin playback of the file before the download is completed.  However, as the HTTP server progressively sends (streams) the whole video content to the client and usually does not take into account how much of the data has been already sent in advance, an abundance of data can overwhelm the video player and lead to a large amount of unused bytes if a user interrupts the video play-out \cite{alcock_application_2011}. To avoid such undesirable situation, the video file is divided into chunks of fixed length and the server pushes them to the client at a rate little higher than the video-bit rate of the transmitted content. As a result, the transmitted traffic creates an ON-OFF pattern, where ON and OFF periods have constant length.

\begin{figure}
\centering
\includegraphics[width=.75\textwidth]{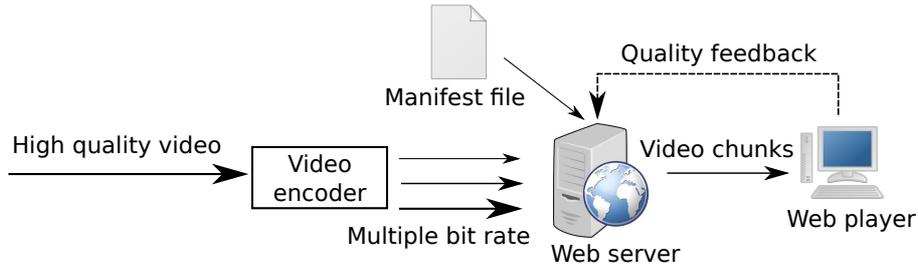}
\caption{Architecture of a video adaptive system based on HTTP}
\label{fig:videoFeedback}
\end{figure}

The extension of this idea is an adaptive streaming, which offers more flexibility when a network environment is less stable, e.g. in wireless mobile networks. With this approach, it is possible to switch the media bit rate (and hence the quality) after each chunk is downloaded and adapt it to the current network conditions \cite{stockhammer_dynamic_2011}. This technique has commenced the development of a new generation of HTTP-based streaming applications which implement client-side play-out algorithms, trying to deliver a continuous stream of video data to end users by mitigating unfavourable network conditions.
In this approach, a video stream is also divided into segments, but this time they are encoded in multiple quality levels, called representations, as drafted in Fig. \ref{fig:videoFeedback}. The algorithm deciding which segment should be requested in order to optimize the viewing experience is a main component and a major challenge in adaptive streaming systems because the client has to properly estimate, and sometimes even predict, network conditions, e.g. the dynamic of available throughput. Furthermore, the client has also to control a filling level of its local buffer in order to avoid underflows resulting in playback interruptions. 

The stream-switching technique is employed today less or more in some proprietary video players, among others in Apple HTTP-based streaming \cite{_http_????}, Microsoft IIS Smooth Streaming \cite{_microsoft_????} or Adobe Dynamic Streaming \cite{_adobe_????}. Moreover, the technique is also adopted by the Dynamic Adaptive Streaming over HTTP (DASH), which is a new MPEG standard pursuing the interoperability between devices and servers of various vendors \cite{sodagar_mpeg-dash_2011}. 

\subsection{Bandwidth estimation algorithm}
As it was already mentioned, the bandwidth estimation algorithm tries to adjust a bit-rate of video to the measured network throughput. As the cited in the section \ref{sec:appFlowControl} adaptive video systems are proprietary and their owners do not give many details about the employed algorithms, we used the adaptive streaming algorithm, with some small modifications, which is implemented in the open-source software described in \cite{muller_vlc_2011}. The approach is relatively simple and its main points were summarised in Algorithm \ref{alg:band}. The examined algorithm calls a function which measures average network throughput $n_b$ in a certain time window $\Delta T$. This window can be considered as a parameter of the algorithm and it may be stretched or shortened in order to optimise streamed video quality. When the video bit-rate $v_b$, which is needed for a smooth video play-out, is lower than the computed average network throughput $n_b$ reduced by $\Delta L$, line \ref{alg:band:upHit}, the algorithm reports that the video quality level $q$ may be increased, i.e. the chunk download module may ask the server for bigger chunks, encoded in higher quality. When the throughput is not sufficient for the given level of video quality, line~\ref{alg:band:downHit}, the opposite situation takes place: the quality level $q$ is decreased and the download module is instructed to obtain chunks of poorer quality what simultaneously demands less network throughput. The parameter $\Delta L$ marks a region of network throughput for which there is no need to switch the quality to a higher level. As a result, the parameter plays a stabilising role and prevents switching the quality levels too frequently, which could have a negative impact on the overall video quality perceived by users. The constants $Q_{\max}$ and $Q_{\min}$ define a range of available levels of the quality.

\renewcommand{\algorithmicrequire}{\textbf{Input:}}
\begin{algorithm}[h]
\begin{center}
\begin{algorithmic}[1]
\REQUIRE $q$ -- current video quality level\\
	$v_b$ -- video bit rate\\
	$\Delta T$ -- time window for measurement of network throughput\\
\STATE $n_b \gets $ getNetworkBandwidth($\Delta T$)
\IF {$v_b < n_b - \Delta L$} 
	\label{alg:band:upHit}
	\IF {$q < Q_{\max}$} \STATE $q \gets q + 1$ \ENDIF
\ENDIF 
\IF {$v_b > n_b$}
	\label{alg:band:downHit}
	\IF {$q > Q_{\min}$} \STATE $q\gets q - 1$ \ENDIF
\ENDIF 
\RETURN $q$
\end{algorithmic}
\end{center}
\caption{Adaptation based on bandwidth estimation}
\label{alg:band}
\end{algorithm}

The rate adaptation algorithm might work fairly well for the case when the player does not share a connection with other flows, network resources are stable and do not fluctuate. Since capacity is measured using an average of recent throughput, the estimate is typically not the same as the true current available capacity. This mismatch results in undesirable behaviour of the streaming algorithms which can be both too conservative and too aggressive. Recent studies have reported many examples of an inaccurate bandwidth estimation while a video client competed against another video client, or against long-lived TCP flow \cite{huang_confused_2012}. In other studies, e.g. in \cite{houdaille_shaping_2012}, it was observed that competing streams behaved instable and unfair among each other what led to significant video quality variation over time. Therefore, some research works, e.g. \cite{zou_can_2015}\cite{bouten_qoe-driven_2015}, try to improve the algorithm by predicting the future bandwidth, while the others, e.g. \cite{huang_using_2014}\cite{huang_downton_2013}\cite{wamser_using_2013}, propose an algorithm based on measurement of buffer occupancy.

\subsection{Buffer reactive algorithm}
The basic idea of the buffer-based reactive algorithm is to select a video bit-rate based on the amount of data that is available in the buffer of a player. Thus, when the buffer reaches a certain level, the system is allowed to increase the quality. Similarly, when draining the buffer, the selected quality is reduced if the buffer shrinks below the threshold. Hence, the quality no longer depends directly on bandwidth availability, and during network outages, buffer under-runs and play-out interruptions may be avoided.

As it was mentioned, the reactive algorithm upgrades the quality once the buffer duration reaches certain chosen thresholds. However with the increasing quality, the video bit-rate rises non-linearly, therefore it is not practical for the buffer threshold to depend directly on the amount of data accumulated in the buffer measured in bytes, but rather on the amount of data measured in video frames, which may be translated into number of seconds of video preloaded in the player buffer. 

Hence, the approach presented in Algorithm \ref{alg:buffer} conserves current video bit rate $q$ as long as the buffer occupancy $b$ remains within the range denoted by $B_{\min}$, line \ref{alg:buffer:upHit} and $B_{\max}$, line \ref{alg:buffer:downHit}. This buffer range plays a role of a cushion which absorbs rate oscillations. If either of these high or low limits are hit, the rate is switched up or down respectively.

\renewcommand{\algorithmicrequire}{\textbf{Input:}}
\begin{algorithm}[h]
\begin{center}
\begin{algorithmic}[1]
	\REQUIRE $b$ -- current buffer occupancy [s]\\
	\label{alg:buffer:bExplained}
\IF {$B_{\max} < b$ } 
\label{alg:buffer:upHit}
	\IF {$q < Q_{\max}$} \STATE $q \gets q + 1$ \ENDIF
\ENDIF 
\IF {$B_{\min} < b $}
\label{alg:buffer:downHit}
	\IF {$q > Q_{\min}$} \STATE $q\gets q - 1$ \ENDIF
\ENDIF 
\RETURN $q$
\end{algorithmic}
\end{center}
\caption{Adaptation based on a level of a player buffer}
\label{alg:buffer}
\end{algorithm}

According to \cite{huang_downton_2013}, buffer reactive algorithms perform fine in many cases, but sometimes they have tendency to too frequent oscillation between video bit rates. Therefore, the authors recommend optimisation of the buffer range confined by $B_{\min}$ and $B_{\max}$ and its adjustment to current network conditions.

\subsection{Hybrid algorithm}
In order to overcome the drawbacks of the solutions presented in Algorithms \ref{alg:band} and \ref{alg:buffer}, we joined their functionality obtaining a hybrid solution, proposed in Algorithm~\ref{alg:hybrid}, which utilises all information available to the bandwidth estimation based and buffer reactive algorithms. When switching the video bit-rate up, the algorithm is cautions and takes into account both the buffer occupancy and the network throughput, refer to the line \ref{alg:hybrid:mainCond}. 

Furthermore, to protect users from video-bit rate oscillations, we added a countermeasure in line~\ref{alg:hybrid:adaptiveCond} in which we check if the number of video-bit rate switches $s$ during the last period $\Delta T$ is within the specified limit $S_{\max}$. When network fluctuations are frequent, the induced by the algorithm changes of the bit-rate have a tendency to cease, making the algorithm insensitive to a variable network environment for time dependent on $S_{\max}$ and $\Delta T$. In order to avoid buffer under-runs, this condition is only verified during the change to a higher video bit-rate.

We assume that the changes of the video bit-rate are counted and updated in a separate code. There also no reasons for both the time windows used for bandwidth measurement and counting of bit-rate switches to be the same length $\Delta T$.

\begin{algorithm}[h]
\begin{center}
	\begin{algorithmic}[1]
\REQUIRE $q$ -- current video quality level\\
	$v_b$ -- video bit rate\\
	$b$ -- current buffer occupancy\\
	$\Delta T$ -- time window for measurement of network throughput\\
	$n_b$ -- measured network bandwidth in the period $\Delta T$\\
	$s$ -- number of video-bit rate switches during the period $\Delta T$
	\label{alg:hybrid:counting}
\IF {$v_b < n_b - \Delta L$ \AND $B_{\max} < b$ } 
\label{alg:hybrid:mainCond}
	\IF {$s < S_{\max}$ } 
	\label{alg:hybrid:adaptiveCond}
		\IF {$q < Q_{\max}$} \STATE $q \gets q + 1$ 
		\ENDIF
	\ENDIF 
\ENDIF

\IF {$v_b > n_b$ \OR $B_{\min} < b$}
	\IF {$q > Q_{\min}$} \STATE $q \gets q - 1$ 
	\ENDIF
\ENDIF 
\RETURN $q$
\end{algorithmic}
\end{center}
\caption{Hybrid adaptation based on bandwidth and buffer occupancy estimation}
\label{alg:hybrid}
\end{algorithm}

\section{Methodology}
\subsection{Quality measures}
\label{ch:qualityMeasures}
From the user's perspective, the key performance characteristic of a network is the QoS of received multimedia content. As the video is transmitted through reliable TCP, no data will be lost. However, there may be play-out interruptions caused by either bandwidth fluctuations or long delays due to retransmissions after packet loss. Furthermore, when reduced network throughput is lower than the playback rate and the buffer will drain, the video playback will pause and wait for new video data. A user expects that delays resulting from content buffering will be minimized and will not occur during a normal video play. Any play-out interruptions are annoying to end users and should be taken into account when estimating the quality of experience (QoE). The QoE is based on popular subjective methods reflecting human perception, as a user is usually not interested in performance measures such as packet loss probability or received throughput, but mainly in the current quality of the received content. However, the quality assessment is time-consuming and cannot be done in real time; therefore, we concentrate on these parameters which we believe impact the QoE at most. We rely, among others, on objective measurement methods, introduced by us in \cite{biernacki_influence_2012}\cite{biernacki_performance_2013}, which for the assessment takes into account video interruptions and its total stalling time.  

The first measurement of the application QoE takes into account relative total stalling time (ST) experienced by a user. Assuming that the video clip is divided into $i$ chunks, each of them has length $\Delta t_i$ we define: 
\begin{equation}
	\text{ST}=\sum_i (\Delta t'_i - \Delta t_i),
\label{eq:SR}
\end{equation}
where $\Delta t'_i$ was the time needed in the reality to play-out the $i$th video chunk and we assume that $t'_i \geq t_i$. It is desirable to minimize the value of the ST by a network operator or a service provider.

\begin{figure}
\centering
\begin{tikzpicture}[scale=.8]
	\draw (0,0) rectangle +(2,1); 
	\draw [dashed] (2,0) --+(1,0);
	\node at (2.4,-.5) {interruption};
	\draw (3,0) rectangle +(2,1); 
	\draw (5,0) rectangle +(2,1); 
	\draw (7,0) rectangle +(2,1); 
	\draw [<->] (9,-2) --+(4,0);
	\node at (11,-1.7) {$\Delta t'_i$};
	\draw [<->] (11,-1) --+(2,0);
	\node at (12,-0.7) {$\Delta t_i$};
	\draw (11,0) rectangle +(2,1);  
	\draw [dotted] (9, 0) -- (9,-2);
	\draw [dotted] (11, 0) -- (11,-1);
	\draw [dotted] (13, 0) -- (13,-2);
	\draw [dashed] (9, 0) -- +(2,0);
\end{tikzpicture}
\caption{Re-buffering events during a play-out of video chunks}
\label{fig:videoBlocks}
\end{figure}
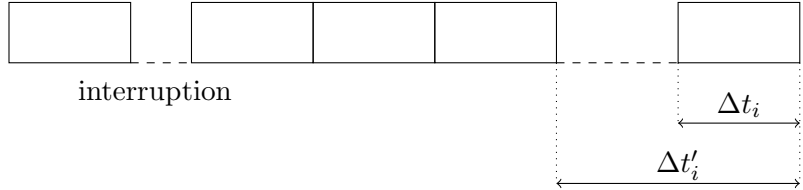

The application QoS defined in Eq. (\ref{eq:SR}) did not differentiate between the cases in which a user can experience one long stalling period or several shorter stalling periods. Thus in our analysis, we also use a second, complementary measurement which quantifies the number of re-buffering events (RE) associated with every stalling period: 
\begin{equation}
	\text{RE}=\sum_i \text{sgn} (\Delta t'_i - \Delta t_i).
\label{eq:RE}
\end{equation}
In practice, if any re-buffering events occur, they will take place before the play-out of the $i$th video chunk when a player waits for its download as presented in Fig. \ref{fig:videoBlocks}.

In Eq. (\ref{eq:RE}), we exclude an initial buffering event which takes place at the beginning of video play, which is used to accommodate throughput variability or inter-packet jitters happening at the beginning of a video play. 

The measurements defined in Eqs. (\ref{eq:SR}) and (\ref{eq:RE}) could quite good characterise performance of non-adaptive video system. However in our case, we can imagine an algorithm that will play-out the stream at the minimum available bit-rate, thus minimising the values of Eqs. (\ref{eq:SR}) and (\ref{eq:RE}), what  will lead to a relatively low video quality and poor utilisation of available network throughput. Hence, we introduce a third measurement which assesses how effectively the algorithm utilises available network resources

\begin{equation}
	\text{TE}=\frac{\sum_i (q_i/Q_i) \Delta t_i}{\sum_i \Delta t_i}.
\label{eq:BE}
\end{equation}

Eq. (\ref{eq:BE}) computes the relation between a quality level $q$ played by an examined algorithm to a theoretical quality level $Q$ which is possible to achieve for given network conditions. The computations take place within discrete time units $\Delta t_i$, into which the video clip is divided, and then are averaged through the duration $\sum_i \Delta t_i$ of the video clip.

The play-out algorithm may try to maximise the measurement presented in Eq.~(\ref{eq:BE}) by adjusting the play-out quality to the given network conditions as frequent as it is possible. Such behaviour will result in rapid oscillations of the video quality, what will be negatively perceived by users \cite{zink_subjective_2003,ni_fine-grained_2009}.
For this reason, we introduce the last measurement, which counts the total number of quality switches (SN) during a video play-out
\begin{equation}
\text{SN}=\sum_i |q_{i+1}-q_{i}|.
\label{eq:SN}
\end{equation}

The design goal of a play-out algorithm is to simultaneously minimise values of the measurements defined in Eqs. (\ref{eq:SR}), (\ref{eq:RE}), (\ref{eq:SN}), and maximise the value of the measurement defined in Eq. (\ref{eq:BE}).

\subsection{Laboratory set-up}
\begin{figure}
\centering
\includegraphics[width=.75\textwidth]{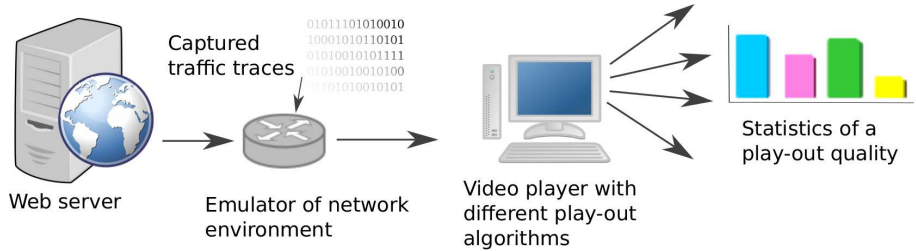}
\caption{Laboratory environment employed for the experiments}
\label{fig:experimentLab}
\end{figure}

In order to capture performance of the adaptive play-out algorithms, we prepared a test environment emulating standard Internet connections encountered in Wi-Fi and HSPA networks. The environment consists of: a web server, video players, a network emulator and a measurement module implemented in the video player, as shown in Fig. \ref{fig:experimentLab}.
The role of the web server plays Apache \cite{_apache_????}, which stores the video clips as a set of chunks. As the video player, we chose VLC media player with the DASH plug-in \cite{muller_vlc_2011}. Both the player and the plug-in have an open-source code, thus it was possible for us to manipulate and completely change the adaptation logic without affecting the other components. As a consequence, the plug-in allowed us to implement and integrate Algorithms \ref{alg:band}, \ref{alg:buffer} and \ref{alg:hybrid} and compare their performance.  
As the network environment model, we used the network emulation node based on the built-in Linux Kernel module \textit{netem} \cite{hemminger_network_2005}. The module is capable of altering network QoS parameters such as network bandwidth or its delays; thus, it allows to test data transmission in different network environments.  

We transmitted several video files, acquired from \cite{lederer_dynamic_2012} and presented in Table~\ref{tbl:videos}, through the simulation environment with variable network throughput. The bandwidth traces were obtained from measurements conducted in WiFi and HSPA networks. For this purpose, we implemented a custom made analysis tool that transferred data at a fixed, configurable rate using UDP packets. Every transmitted packet had its sequence number, enabling the receiver to precisely detect packet loss, and a time-stamp showing when the package left the receiver. The packets were sent at a fixed rate, and the packet reception rate was logged. The tests measured the download performance as it seems to be more important than upload performance for a one-way video streaming scenario.

\begin{table}
\begin{center}
\caption{Video clips used during the experiments}
\label{tbl:videos}
\small
\begin{tabular}{|l|l|p{4cm}|}
\hline\noalign{\smallskip}
Name & Genre& Bitrate levels\\
\noalign{\smallskip}\hline\noalign{\smallskip}
Big Buck Bunny & animation & \multirow{6}{*}{\vbox{150~kbit/s -- 320x240, 
						300~kbit/s -- 480x360,
						600~kbit/s -- 854x480,
						1.2~Mbit/s -- 1280x720,
						2.5~Mbit/s -- 1920x1080}}\\
Elephants Dream&animation&\\
Red Bull Playstreets &sport&\\
The Swiss Account&sport&\\
Valkaama&movie&\\
Of Forest and Men&movie&\\
\noalign{\smallskip}\hline
\end{tabular}
\end{center}
\end{table}

The captured log was used as a template for the bandwidth shaper implemented in the mentioned earlier \textit{netem} module. In addition to bandwidth throttling, the \textit{netem} module also adds a delay of 20~ms to WI-FI connection and 100~ms to HSPA connection to emulate the average latency which was experienced and measured during gathering of the throughput traces. In order to obtain desirable, average network throughputs, which were used in the experiments, the traces were rescaled. Thus, having identical bandwidth trace we were able to perform a quite fair and realistic comparison of the play-out algorithms.

We believe that the above described methodology provides an attractive middle ground between simulation and real network experiments. To a large degree, the emulator should be able to maintain the repeatability, reconfigurability, isolation from production networks, and manageability of simulation while preserving the support for real video adaptive applications. 

Using our laboratory environment, we compared the three presented solutions with the parameters specified in Table~\ref{tbl:algParameters}. The hybrid solution was applied in two versions: the base one, which tries to take advantage of network conditions in order to increase video quality without taking into account the buffer state of a player; and the adaptive one, which is more conservative and increases video quality after taking not only the network state and the buffer state, but also frequency of previous bit-rate switches. Each compared algorithm played first 600~s of every video clips presented in Table \ref{tbl:videos}.

\begin{table}
\begin{center}
\caption{Algorithms and their parameters used in the experiments}
\label{tbl:algParameters}
\small
\begin{tabular}{|l|p{7cm}|}
\hline\noalign{\smallskip}
Algorithm & Parameters\\
\noalign{\smallskip}\hline\noalign{\smallskip}
Bandwidth est. (Alg. \ref{alg:band})&$\Delta T=4 s$, $\Delta L=0.25n_b$\\
Buffer reactive (Alg. \ref{alg:buffer})&$B_{\min}=3 s$, $B_{\max}=7 s$\\
Hybrid, basic (Alg. \ref{alg:hybrid}) &$\Delta T$, $\Delta L$, $B_{\min}$ -- as above, $\Delta T_s=10s$, $B_{\max}=0 s$, $S_{\max}=\inf$\\
Hybrid, adaptive (Alg. \ref{alg:hybrid})&$\Delta T$, $\Delta L$, $B_{\min}$, $B_{\max}$, $\Delta T_s$ -- as above, $B_{\max}=7 s$, $S_{\max}=10$\\
\noalign{\smallskip}\hline
\end{tabular}
\end{center}
\end{table}

\section{Results}
\begin{figure}
\centering
\includegraphics[width=.75\textwidth]{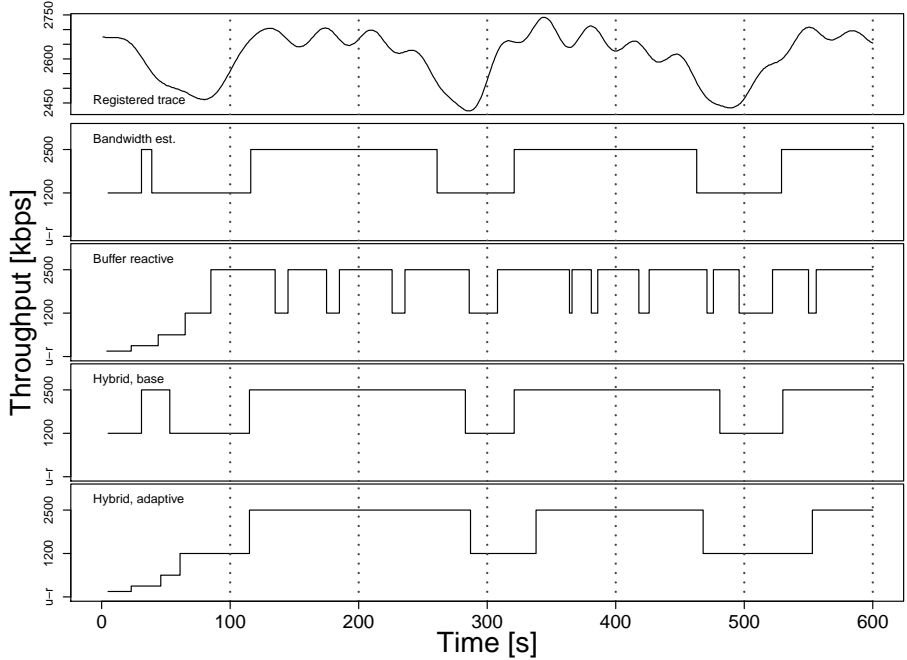}
\caption{Transient comparison of the play-out algorithms in an emulated Wi-Fi environment. Average bandwidth set to 2600~kbps, video clip Big Buck Bunny (see Table \ref{tbl:videos})}
\label{fig:constBandMergedTrace}
\end{figure}

The output of the experiments performed in the Wi-Fi environment is presented in Fig. \ref{fig:constBandMergedTrace}. The network throughput oscillates between about 2450~kbps and 2700~kbps with an average set to 2600~kbps; however, one must notice that due to TCP/IP and other protocols overhead the effective throughout available for the video streaming is a few percent lower.

As the effective network throughput fluctuates near the highest available in the experiment video bit-rate, the play-out algorithm based on bandwidth estimation quite regularly switches between 1200~kbps and 2500~kbps. The switches are correlated with the local minima of the network throughput.

The buffer reactive algorithm starts from the lowest available bit rate and, as the buffer is being filled with data, gradually increases the quality, reaching 2500~kbps in about 80th~s of the playback. Then, the playback quality starts to switch between 2500~kbps and 1200~kbps; nevertheless, the oscillations are less regular and more frequent compared to the bandwidth estimation algorithm.

Because the hybrid algorithm on the beginning of the play-out measures available throughput, it is able to start with a higher quality level compared to its buffer reactive competitor. Furthermore, the algorithm loses the highest quality rate less frequently, mostly in the times, when the available network throughput achieves local minima. A visual assessment may lead to a conclusion that in this particular experiment, the hybrid algorithm obtains better performance than its competitors, at least taking into account bandwidth effectiveness, which was defined in Eq. (\ref{eq:BE}).

The adaptive version of the hybrid algorithm starts its playback similarly to the buffer reactive algorithm: from the lowest quality, then gradually reaching the maximum. Because of this gradual improvement of quality, the algorithm is able to maintain the streaming at 2500~kbps a bit longer compared to its base version. Additionally, the algorithm spends more time streaming 1200~kbps video compared to its base version, which is a result of the condition defined in line \ref{alg:hybrid:mainCond} of Algorithm \ref{alg:hybrid}, which takes into account both network bandwidth and the buffer state of a player into consideration when switching the quality rate to higher level. This leads to a drop in throughput efficiency of the algorithm, while the number of bit-rate switches is comparable to the base version of the algorithm.\\

The same experiments were performed for the rest of the movies listed in Table \ref{tbl:videos} and were extended to the cases, where the network throughput was set in average to \thrLow~and \thrMed. These values were chosen to a certain extent arbitrarily, however we took into account that in the first case, the effective network throughput falls in the middle of 600~kbps and 1200~kbps levels of the available video bit-rate; and in the second case, the effective network throughput is theoretically a little above the fifth defined quality level of 1200~kbps.

\begin{figure}
\centering
	\subfigure[Delay]{\label{fig:bandConstD} \includegraphics[width=\multiPicWidth\linewidth]{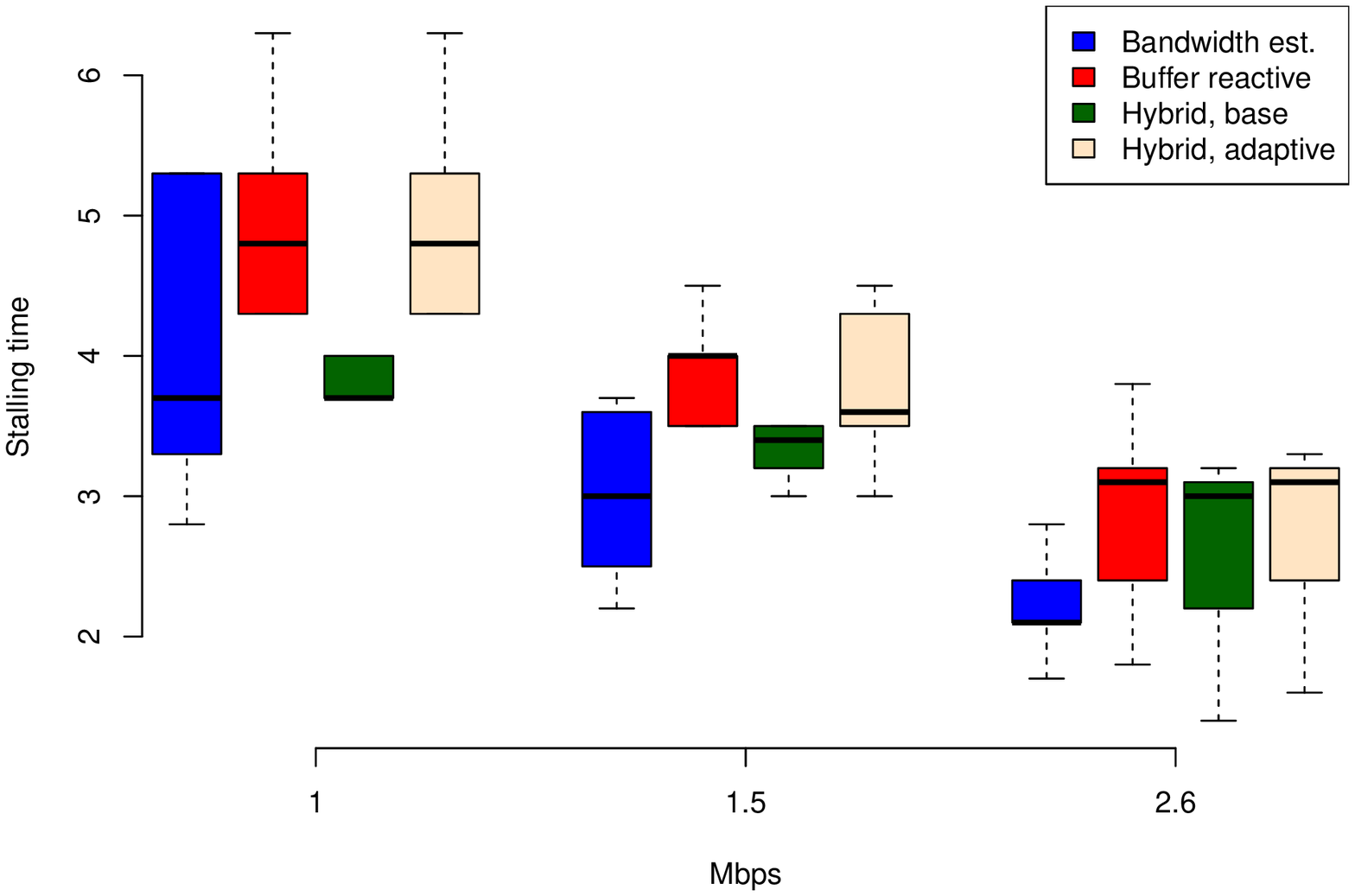}}
        \subfigure[Stalling events]{\label{fig:bandConstS} \includegraphics[width=\multiPicWidth\linewidth]{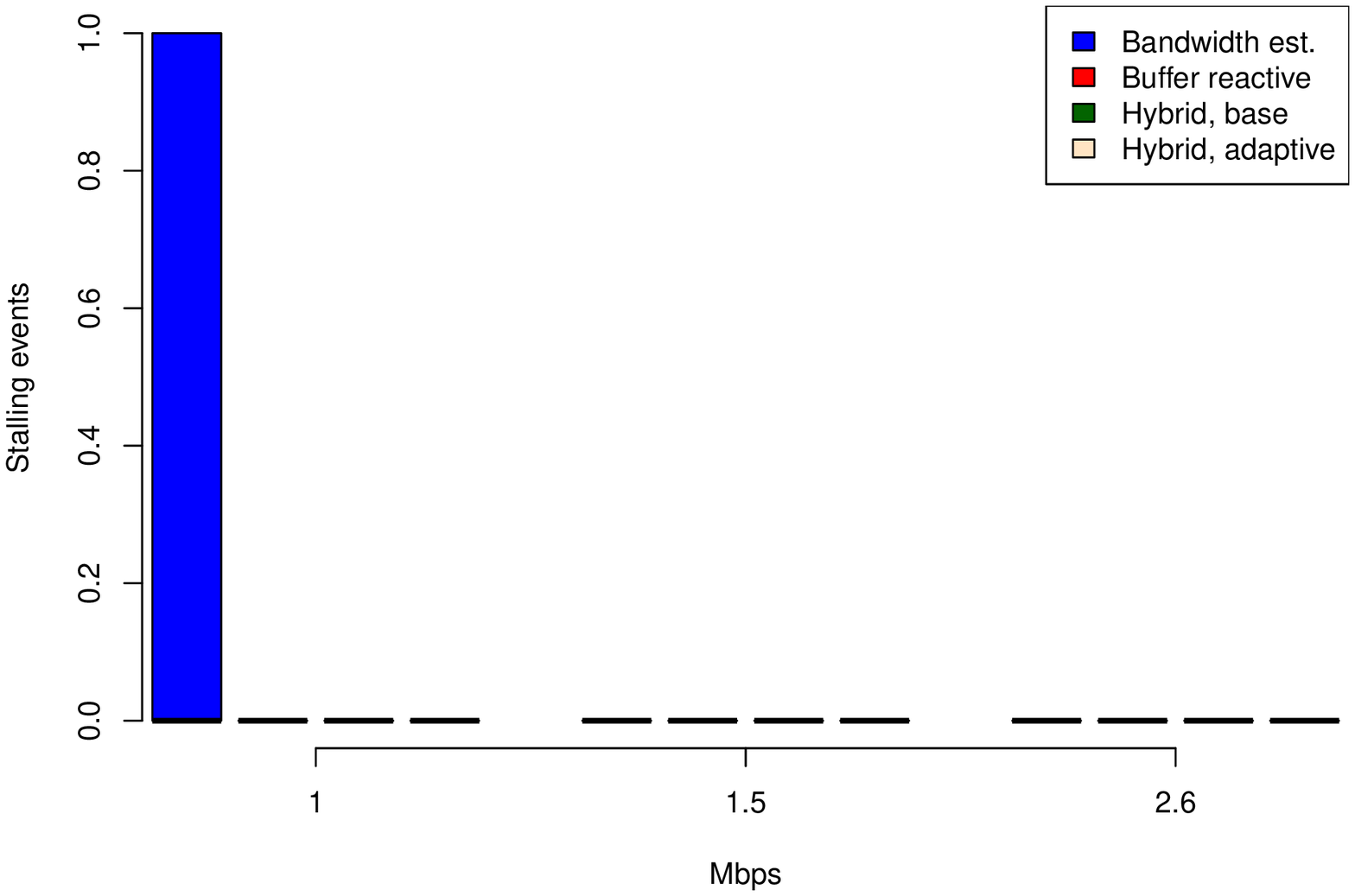}}
        \\
	\subfigure[Efficiency]{\label{fig:bandConstE} \includegraphics[width=\multiPicWidth\linewidth]{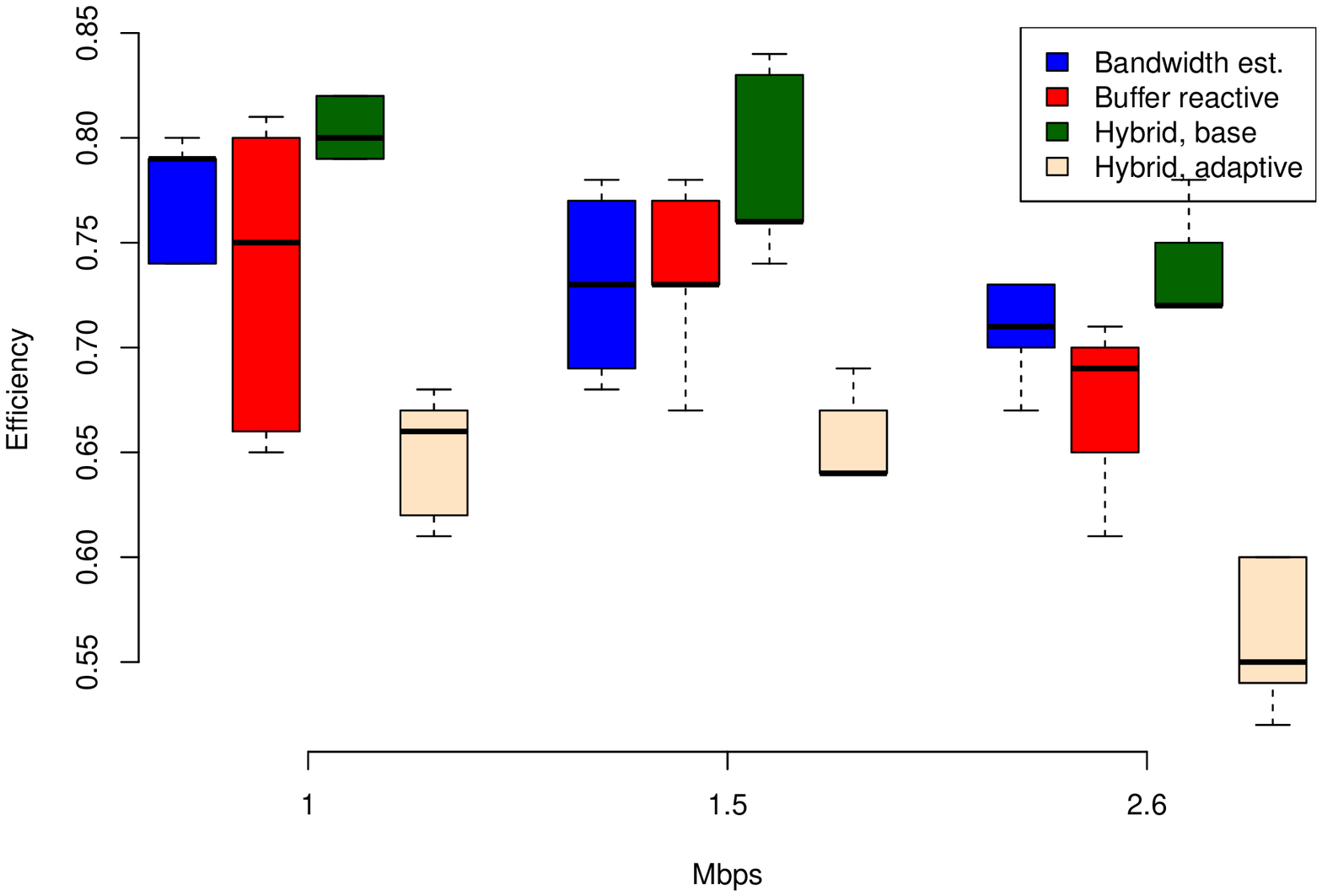}}
	\subfigure[Number of switches]{\label{fig:bandConstN} \includegraphics[width=\multiPicWidth\linewidth]{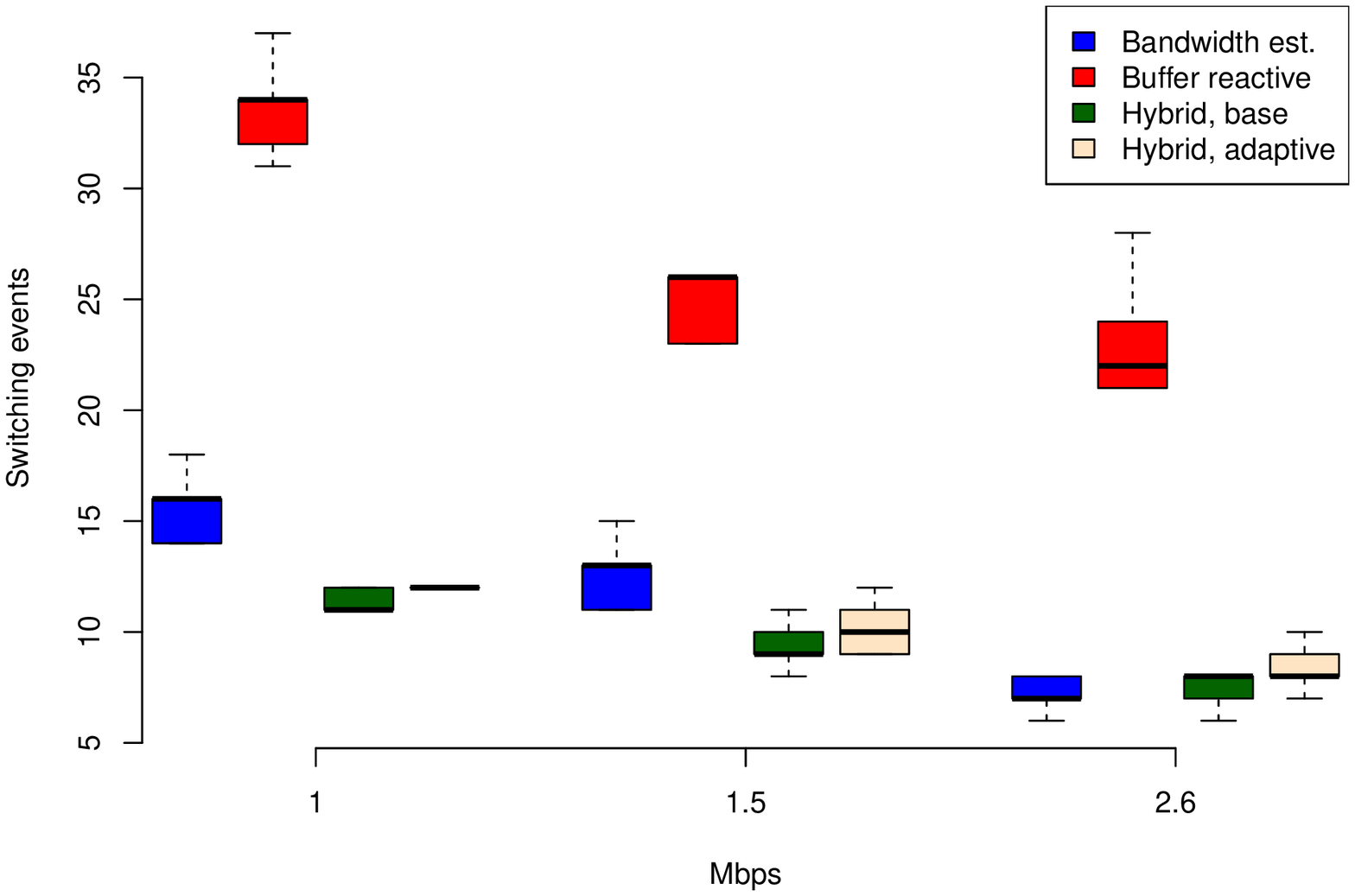}}
                \caption{Comparison of the play-out algorithms in an emulated Wi-Fi environment}
        \label{fig:bandConst}
\end{figure}

The averages of the relative stalling time (ST), defined in Eq. (\ref{eq:SR}), are similar for all examined algorithms, as it was presented in Fig. \ref{fig:bandConstD}. For the throughput set to \thrLow, the average ST is between about 3~s and 5~s. With an increase of the throughput, the average ST decreases, taking values between about 3~s and 4~s for the throughput set to \thrMed, and between about 2~s and 3.5~s for the throughput set to \thrHigh. The main observable difference among the algorithms is that the bandwidth estimation approach has a little higher variation compared to the other solutions, especially for throughput set to \thrLow. 

The higher variability of the bit-rate in the case of the bandwidth estimation algorithm may be explained when we examine the number of re-buffering events (RE) defined in Eq.~(\ref{eq:RE}) and presented in Fig. \ref{fig:bandConstS}. During the video streaming at the lowest network throughput, the algorithm experienced at least once a re-buffering event, which influenced also the ST. The re-buffering was probably caused by an overlap of two unfavourable factors: an overoptimistic estimation of the network throughput and a subsequent burst of the bit-rate in the transmitted variable bit-rate video stream. Other algorithms did not experience buffer under-runs.

The average throughput efficiency, defined in Eq. (\ref{eq:BE}), for all algorithms but the adaptive version of the hybrid are comparable, and in most cases ranges between 65\% and 80\%, as it was presented in Fig. \ref{fig:bandConstE}. As expected from the transient analysis of quality traces presented in Fig. \ref{fig:constBandMergedTrace}, the TE for the adaptive hybrid solution is clearly lower, achieving about 65\% for the average throughput set to \thrHigh~or \thrMed, and less than 60\% for the throughput set to \thrLow.

The frequency of the bit-rate switching is the highest for the buffer reactive algorithm: from more than 30 switches for the throughput set to \thrLow~to about 25 switches for the throughput set to \thrMed~and \thrHigh, as shown in Fig.~\ref{fig:bandConstN}. However, we must note that the cautious increase of the bit-rate in the first minute of the play-out, see Fig.~\ref{fig:constBandMergedTrace}, influences negatively the algorithm score. The bandwidth reactive solution achieves significantly better results, ranging from about 15 switches for \thrLow~network throughput, through about 12 for \thrMed, and less than 10 for \thrHigh~throughput. The hybrid approaches achieve the best results, experiencing between 6 and 12 switches in the case of the base version of the algorithms, and from 7 to 13 switches for the adaptive version of the algorithm. Simultaneously, these scores have the lowest variation. Similarly to the buffer reactive solution, the score of the adaptive version of the hybrid approach is negatively biased due to the gradual increase of the video bit-rate on the beginning of the play-out.\\

The next series of the experiments were performed in an emulated mobile network environment based on captured traces from an HSPA system. Similarly to the  Wi-Fi scenarios, on the beginning we examine the behaviour of the play-out algorithm for a network with an average throughput set to 2600~kbps; however, with much wider range of its oscillations, stretching from a level where the network totally collapses (being the result of e.g. signal fading) and throughput drops to zero~kbps, to situations where the throughput is higher than 5000~kbps. 

The beginning of the play-out, the network throughput is relatively high, therefore, the algorithm based on bandwidth estimation is able to serve video above its average quality. Nevertheless, with deteriorating network conditions, the algorithm rapidly decreases the transmission quality, what quite often ends up with congestions -- undesired condition not met during the experiments in the Wi-Fi network. After the network collapses, its throughput has a tendency to shoot up, what in consequence translates to a rapid increase of the play-out quality. As a result, the quality record has a fairly high range of oscillations in short periods of time what may negatively influence end users' perception.

Similarly to the behaviour of the algorithm based on bandwidth estimation, the buffer reactive algorithm also reacts nervously to the rapid throughput fluctuations. The video quality jumps several times from 1200~kbps onto 2500~kbps, but the algorithm usually persists in serving the video at the highest level only for a relatively short period of time. Contrary to the previous algorithm, the buffer reactive solution handles better network collapses. The first and second network slips, which happen in about 75th~s and 175th~s respectively, remain almost unnoticed to users. In other similar critical situations, the algorithm struggles to keep up continuity of the play-out, however at the cost of reduction of its bit-rate to the lowest possible level. Generally, compared to the bandwidth estimation approach, the oscillation range of the video quality is a bit tighter, although the rate of the quality switches remains higher.

Compared to its two predecessors, the hybrid algorithm responds more calmly to the throughput fluctuation. The algorithm is able to maintain the highest quality level longer and avoids so frequent quality switches as the buffer reactive algorithm. Simultaneously, the hybrid solution handles better the throughput falls than the play-out regulated by the bandwidth estimation technique, being able to escape from stalls during the streaming and sudden plunges of the video quality.    

The introduction of additional hedging against frequent bit-rate switches leads to a smoother quality trace, although we may still observe a few needless tries of quality improvement. The smoother quality comes at a certain price: it is traded for a lower average bit-rate what translates to significantly poorer efficiency of network throughput usage.\\

\begin{figure}
\centering
\includegraphics[width=.75\textwidth]{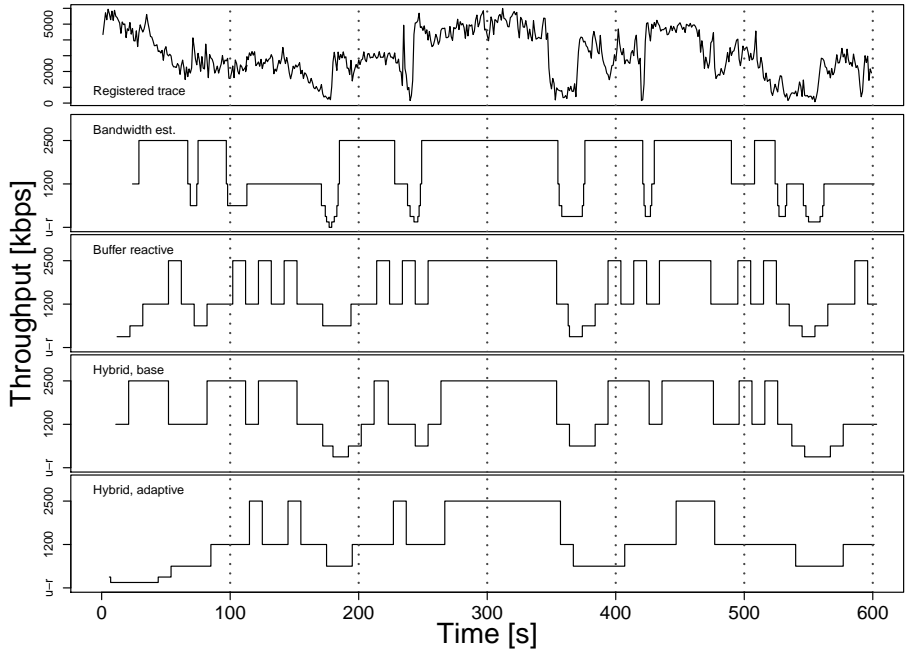}
\caption{Transient comparison of the play-out algorithms in an emulated wireless mobile environment. Average bandwidth set to 2600~kbps, video clip Big Buck Bunny (see Table \ref{tbl:videos})}
\label{fig:varBandMergedTrace}
\end{figure}

\begin{figure}
\centering
	\subfigure[Delay]{\label{fig:bandVarD} \includegraphics[width=\multiPicWidth\linewidth]{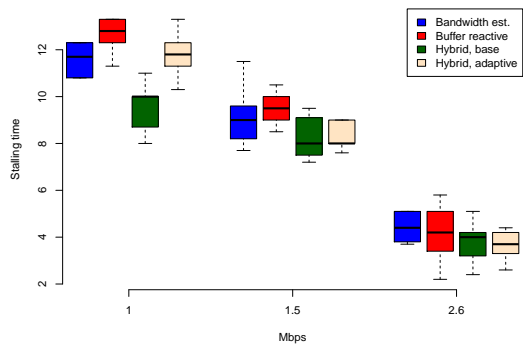}}
        \subfigure[Stalling events]{\label{fig:bandVarS} \includegraphics[width=\multiPicWidth\linewidth]{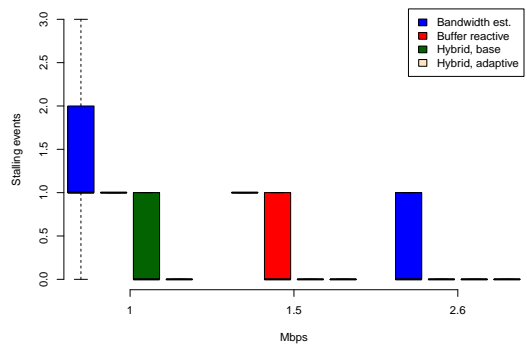}}
        \\
	\subfigure[Efficiency]{\label{fig:bandVarE} \includegraphics[width=\multiPicWidth\linewidth]{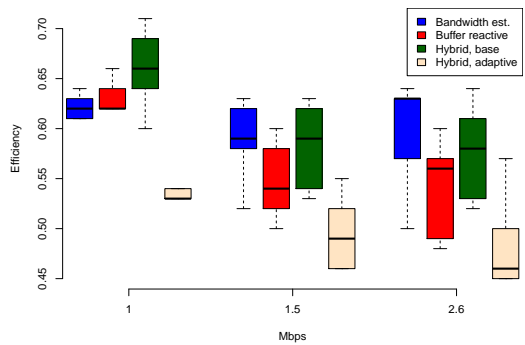}}
	\subfigure[Number of switches]{\label{fig:bandVarN} \includegraphics[width=\multiPicWidth\linewidth]{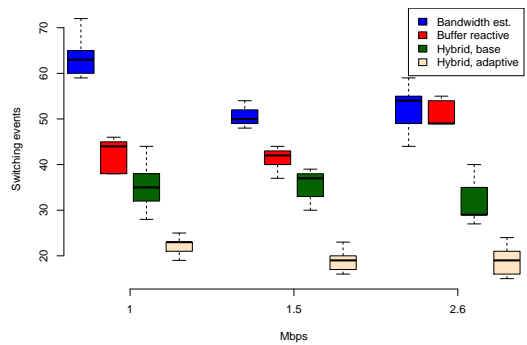}}
                \caption{Comparison of the play-out algorithms in an emulated wireless mobile environment}
        \label{fig:bandVar}
\end{figure}

Similarly to the examination of the algorithms in the Wi-Fi environment, we extended our analysis to the rest of the movies from Table \ref{tbl:videos}. The network throughput was set in average to \thrLow, \thrMed~and \thrHigh, which are the same values as in the case of the Wi-Fi experiment, in order to make the comparison of the algorithms easier in these two environments.  

The ST is roughly comparable for all algorithms, as it was pictured in Fig. \ref{fig:bandVarD}. There are some noticeable differences for the throughput set to \thrLow; although, after closer examination, their absolute values are of about several seconds what should not have much influence on end users' experience. As expected, with the increasing network throughput, the ST decreases, dropping from about 9~s - 13~s in the case of \thrLow~throughput, to about 3~s - 5~s in the case of \thrHigh~throughput.   

For the average throughput set to \thrLow, the bandwidth assessment algorithm experiences up to 3 stalling events, depending on the video clip played. The buffer reactive algorithm experiences in average 1 stalling event.  The hybrid solution in its base form experiences up to 1 stalling event while its adaptive version is free from stallings. With the increasing throughput, the probability of stalling is lower, however the bandwidth estimation approach still has some problems with a smooth play-out. Also for the throughput set to \thrMed, the buffer reactive algorithm sporadically experiences breaks during the play-out. In contrast, both hybrid solutions are able to deliver the video without interruptions.

When it comes to the assessment of throughput efficiency, the base version of the hybrid algorithm clearly outperforms the rest when the throughput is set to \thrLow, as shown in Fig. \ref{fig:bandVarE}. When the throughput increases, the hybrid solution loses its advantage over the competitors. The buffer assessment approach notes slightly better result for throughput set to \thrLow~compared to the bandwidth one; nonetheless, when the throughput rises to \thrMed~or \thrHigh, its efficiency drops below the efficiency of the bandwidth assessment approach. The adaptive version of the hybrid algorithm obtains the lowest efficiency from all compared solutions. 
Because the mobile network has more dynamical fluctuation of its throughput, the efficiency of the algorithms operating in this environment is in general about 20\% worse compared to these examined in the Wi-Fi environment, see Fig. \ref{fig:bandConstE}. 

The bandwidth assessment approach clearly under-performs in the SN experienced during the play-out. For the throughput set to \thrLow, the SN reaches nearly 70, which is about 30\% higher than the second score of 45 switches for the buffer reactive algorithm. The hybrid solution achieves less than 45 switches for its base version and less than 25 switches for its adaptive version. The increase of the network throughput to \thrMed~reduces the SN for the bandwidth algorithm to about 50 switches, which is however still not enough to outperform the buffer reactive and hybrid algorithms, which achieve even better results than in the case of \thrLow~throughput. Further rise of the throughput to \thrHigh, aligns the results for bandwidth and buffer reactive algorithms to about 50 switches, while the hybrid algorithms still achieve significantly better score. 

The evaluation shows that the hybrid approach in most cases achieves better results compared to its competitors. 
In the Wi-Fi environment, where the network conditions are relatively stable and the throughput fluctuation has relatively low amplitude, the differences between the examined algorithms are mainly visible when we take into account efficiency of utilisation of network throughput and frequency of video bit-rate switches. The base hybrid strategy obtained in some scenarios about about 10\% better throughput utilisation compared to the solutions based on buffer or bandwidth assessment, thereby it is able to play video of better quality. Simultaneously, the base hybrid solution obtains no worse results than its competitors in other performance measurements. 

When we take into account the mobile network, the hybrid strategy is even more dominant. Except achieving significantly lower switching of the video-bit rate, the hybrid approach is free from under-runs of a player buffer which causes a video clip to stall in a middle of a play-out, what takes place in the case of two other strategies. Compared to the aggressive hybrid strategy, its adaptive version achieves usually better results when taking into account the stability of the play-out, especially in the mobile environment. However, this score is traded for significantly worse throughput utilisation compared to other strategies.

\section{Conclusions}
In the paper we proposed a novel algorithm dedicated for an adaptive streaming based on HTTP. The algorithm employs a hybrid play-out strategy which combines two popular approaches: a bandwidth estimation and a buffer control. As a consequence, we assumed that the hybrid solution will exploit strengths of both approaches and it will avoid their weaknesses. The proposed algorithm was implemented in two versions which differ in the method of handling throughput fluctuations. The first, base version tries to aggressively exploit any favourable network conditions in order to increase the bit-rate of played video. The second version is equipped with an adaptive accent: it increases the streamed bit-rate more carefully, taking into account the buffer state of a player and throughput variability.

The evaluation shows that in the mobile networks, where network throughput has lots of variability, the hybrid approach achieves better performance compared to traditional solutions based on bandwidth estimation or buffer assessment. The advantages of hybrid solution are less visible in networks where changes in network throughput have lower amplitude. Such a result is a consequence of the construction of the examined algorithms. Single parameters which describe video systems, like network throughput or amount of buffered video, tend to strictly depend on network conditions. The more variable the conditions are, the more variable the parameters. When the algorithms take into account only a one of these parameters, naturally the quality of the transmission will strictly depend on the current network condition described by this parameter. Therefore, an introduction of additional components which the algorithm takes into account when making decisions about a bit-rate adaptation, stabilises and improves a quality of a play-out. In the proposed solution, the algorithm considers two components and gives them equal weight. However, one can study scenarios, with multiple components describing the state of a video system, which may not only include network throughput or buffer level, but also their pace of change, deviations, averages etc. Some of these components may have priorities, which may be dependent on the characteristic of a network environment.  

\bibliographystyle{unsrt}
\bibliography{bib}

 \end{document}